\documentclass[12pt]{article}
\usepackage{amssymb}
\usepackage{amsbsy}
\usepackage{epsfig}
\usepackage{verbatim}
\makeatletter \@addtoreset{figure}{section}
\def\thefigure{\thesection.\@arabic\c@figure}
\def\fps@figure{h, t}
\@addtoreset{table}{bsection}
\def\thetable{\thesection.\@arabic\c@table}
\def\fps@table{h, t}
\@addtoreset{equation}{section}

\newtheorem{corollary}{Corollary}[section]
\newtheorem{definition}{Definition}[section]
\newtheorem{theorem}{Theorem}[section]
\newtheorem{proposition}{Proposition}[section] 

\newtheorem{examps}{Examples}[section]

\newtheorem{lemma}{Lemma}[section]
\newtheorem{remark}{Remark}[section]
\newtheorem{remarks}[remark]{Remarks}

\def\bx{\begin{example}}
\def\ex{\end{example}}
\def\bxs{\begin{examps}. \rm\begin{enumerate}}
\def\exs{\end{enumerate}\end{examps}}
\def\bd{\begin{definition}}
\def\ed{\end{definition}}
\def\bt{\begin{theorem}}
\def\et{\end{theorem}}
\def\bp{\begin{proposition}\rm}
\def\ep{\end{proposition}}
\def\bc{\begin{corollary}}
\def\ec{\end{corollary}}
\def\bl{\begin{lemma}\em}
\def\el{\end{lemma}}
\def\be{\begin{equation}}
\def\ee{\end{equation}}
\def\br{\begin{remark}\rm\small}
\def\er{\end{remark}}
\def\brs{\begin{remarks}.\\ \rm\
\begin{enumerate}}
\def\ers{\end{enumerate}\end{remarks}}
\def\bea{\begin{eqnarray}}
\def\eea{\end{eqnarray}}

\def\ra{{\rightarrow}}

\def\wt{\widetilde}

\def\det{\mathrm {det}}
\def\sgn{\mathrm {sgn}}

\def\diag{\mathrm {diag}}

\def\ds{\displaystyle}

\def\&{&{\hskip -20pt}}

\def\HH{{\mathcal H}}
\def\KK{{\mathcal K}}

\def\PP{{\mathcal P}}

\def\Ib{{\mathbf I}}

\def\Nb{{\mathbf N}}

\date{}

\begin{document}
\baselineskip 16pt 
\begin{flushright}
CRM-3196 (2005)
\end{flushright}
\medskip
\begin{center}
\begin{Large}\fontfamily{cmss}
\fontsize{17pt}{27pt}
\selectfont
\textbf{Integrals of Rational Symmetric Functions, \\ 
Two--Matrix Models and Biorthogonal Polynomials}\footnote{
Work supported in part by the Natural Sciences and Engineering Research Council
of Canada (NSERC) and the Fonds FCAR du Qu\'ebec.}
\end{Large}\\
\bigskip
\begin{large}  {J. Harnad}$^{\dagger \ddagger}$\footnote{harnad@crm.umontreal.ca}
 and {A. Yu. Orlov}$^{\star}$\footnote{ orlovs@wave.sio.rssi.ru}
\end{large}
\\
\bigskip
\begin{small}
$^{\dagger}$ {\em Centre de recherches math\'ematiques,
Universit\'e de Montr\'eal\\ C.~P.~6128, succ. centre ville, Montr\'eal,
Qu\'ebec, Canada H3C 3J7} \\
\smallskip
$^{\ddagger}$ {\em Department of Mathematics and
Statistics, Concordia University\\ 7141 Sherbrooke W., Montr\'eal, Qu\'ebec,
Canada H4B 1R6} \\ 
\smallskip
$^{\star}$ {\em Nonlinear Wave Processes Laboratory, \\
Oceanology Institute, 36 Nakhimovskii Prospect\\
Moscow 117851, Russia } \\
\end{small}
\end{center}
\bigskip
\bigskip
\begin{center}{\bf Abstract}
\end{center}
\smallskip

\begin{small}
We give a new method for the evaluation of a class of integrals of rational symmetric functions in $N$ pairs of variables $\{(x_a, y_a)\}_{a=1,\dots N}$  arising in coupled matrix models, valid for a broad class of two-variable measures. The result is expressed as the determinant of a matrix whose entries consist of the associated biorthogonal polynomials, their Hilbert transforms, evaluated at the zeros and poles of the integrand, and bilinear expressions in these. The method is elementary and direct, using only standard determinantal identities, partial fraction expansions and the property of biorthogonality. The corresponding result for one-matrix models and integrals of
rational symmetric functions in $N$ variables $\{x_a\}_{a=1,\dots N}$ is also rederived in
a simple way using this method.

\bigskip
\end{small}
\bigskip \bigskip

\section{Integrals of symmetric rational functions}

Let $d\mu(x,y)$ be a measure (in general, complex), supported on a finite set of products of curves in the complex $x$ and $y$ planes, for which the semi-infinite matrix of bi-moments is finite:
\be
B_{jk} := \int d\mu(x,y)x^j y^k <\infty, \quad 0 , \quad \forall j,k \in {\bf N}.
\ee
The integrals are understood to be evaluated on a specified linear combination of products of the support curves. Assuming that, for all $N\ge 1$, the $N \times N$ submatrix  $(B_{jk})_{0\le j,k, \le N-1}$ is nonsingular, the Gram-Schmidt process may be used to construct an infinite sequence of pairs of biorthogonal polynomials $\{P_j (x), S_j(y)\}_{j=0\dots \infty}$, unique up to signs, satisfying 
\be
\int d\mu(x,y) P_j(x)S_k(y) = \delta_{jk},
\ee
and normalized to have leading coefficients that are  equal:
\be
P_j(x) = {x^j\over\sqrt{h_j}} + O(x^{j-1}), \qquad S_j(x) = {y^j\over \sqrt{h_j} }+ O(y^{j-1}).
\ee
We will also assume that  the Hilbert transforms of these biorthogonal polynomials,
\be
\tilde{P}_n(\mu) := \int d\mu(x,y) {P_m(x) \over \mu -y},  \quad
\tilde{S}_n(\eta) := \int d\mu(x,y) {S_m(x) \over \eta -x},
\ee
exist for all $n\in {\bf N}$.

 For $N\ge 1$, let
\be
{\bf Z}^{(2)}_N := \int d\mu(x_1, y_1) \dots \int d\mu(x_N, y_N) \Delta_N(x) \Delta(y)
=N! \prod_{n=0}^N h_n
 \label{partfn}
\ee
where
\be
\Delta_N(x) := \prod_{i > j}^N (x_i-x_j), \quad \Delta_N(y) := \prod_{i > j}^N (y_i-y_j)
\ee
are Vandermonde determinants. Such integrals are of particular interest in two-matrix models \cite{BEH1, BEH2, EM, GMO, IZ, HO1} and may be interpreted as the reduction to the space of eigenvalues of  the integral defining the partition function on an ensemble of pairs of $N \times N$ matrices having a probability measure that is invariant with respect to conjugation by  unitary matrices, and admitting a Harish-Chandra-Itzykson-Zuber \cite{IZ} type reduction to the $2N$ dimensional space of eigenvalues. The joint probability distribution (in the case of a real Borel measure) is given, after normalization,  by the integrand of (\ref{partfn}) with eigenvalues $(x_1, \dots x_N), (y_1, \dots y_N)$ having values on the support curves of the measure $d\mu(x,y)$.  

Now choose four sets of complex constants 
$\{\xi_\alpha, \zeta_\beta,\eta_j, \mu_k\}_{\alpha =1, \dots L_1, \beta = 1, \dots L_2 \atop j =1 \dots M_1, k = 1 \dots M_2}$, distinct within each of the four groups, with
\bea
N+L_1-M_1 &\&\ge N+ L_2 - M_2 \ge 0
\label{LNMinequal}
\eea
and such that the $\eta$'s and $\mu$'s are not on the support curves of the
bimeasure $d \mu(x,y)$  in the $x$ and $y$ planes, respectively. (There is no loss of 
generality in assuming the first inequality in (\ref{LNMinequal}), since the results for the reversed  case can be read off by symmetry in the two sets of variables.)
The  main result of this work is a new derivation of the following expression for the integral of a symmetric, rational function in two sets of $N$ variables $\{x_a , y_a\}_{a=1,\dots N}$, having simple zeros at the points  $\{\xi_\alpha\}_{\alpha = 1 \dots L_1}$, $\{\zeta_\beta\}_{\beta = 1 \dots L_2}$  and simple poles at $\{\eta_j\}_{j= 1 \dots M_1}$, $\{\mu_k\}_{k = 1 \dots M_2}$ in terms of the  biorthogonal 
polynomials $\{P_n, S_n\}$, evaluated at the points $\{\xi_\alpha\}_{\alpha = 1 \dots L_1}$, $\{\zeta_\beta\}_{\beta = 1 \dots L_2}$, and bilinear combinations of these,  together with the Hilbert transforms $\{\tilde{P}_n, \tilde{S}_n\}$ evaluated at the points $\{\mu_k\}_{k = 1 \dots M_2}$, $\{\eta_j\}_{j= 1 \dots M_1}$.
\bea
 {\bf I}^{(2)}_N &:=& {1\over {\bf Z}^{(2)}_N} \int d\mu(x_1,y_1) \dots \int d\mu(x_N,y_N)
 \Delta_N(x)\Delta_N(y) \cr
 && \quad  \times \prod_{a=1}^N {\prod_{\alpha=1}^{L_1}(\xi_\alpha - x_a) \prod_{\beta=1}^{L_2}(\zeta_\beta - y_a) \over \prod_{j=1}^{M_1}(\eta_j - x_a)  \prod_{k=1}^{M_2}(\mu_k - y_a)}
   \label{I2Ndef} \\
 &=&   \epsilon(L_1,L_2,M_2,M_2)
 \prod_{n=N}^{{}_{N+L_2-M_2-1}}{\hskip -15 pt} \sqrt{h_n} \prod_{n=N}^{{}_{N+L_1- M_1-1}}{\hskip -15 pt}\sqrt{h_n} \cr
 &\& \quad \times
{ \prod_{\alpha=1}^{L_1}\prod_{j=1}^{M_1}(\xi_\alpha - \eta_j)\prod_{\beta=1}^{L_2}\prod_{k=1}^{M_2}(\zeta_\beta- \mu_k)
 \over \Delta_{L_1}(\xi)  \Delta_{L_2}(\zeta)  \Delta_{M_1}(\eta)  \Delta_{M_2}(\mu)}
 \  \det G,  \cr
 &&
 \label{mainintegral}
\eea
where 
\be
\epsilon(L_1,L_2,M_2,M_2):=  (-1)^{{1\over 2}(M_1+M_2)(M_1+M_2-1)}(-1)^{L_1 M_2}
\label{epsilon1234}
 \ee
 and $G$ is the $(L_1 + M_2) \times (L_1 + M_2) $ matrix 
\be
G= \pmatrix{ \ds{\mathop{K_{11}}^{\!N+L_2-M_2}}(\xi_\alpha, \eta_j) & \ds{\mathop{K_{12}}^{\!N+L_2-M_2}}(\xi_\alpha, \zeta_\beta) & 
P_{N+L_2 - M_2}(\xi_\alpha)  & \dots & P_{N+L_1 - M_1-1}(\xi_\alpha) \cr
\ds{\mathop{
K_{21}}^{\!N+L_2 - M_2}}(\mu_k, \eta_j) & \ds{\mathop{ K_{22}}^{\!N+L_2-M_2}}(\mu_k, \zeta_\beta) & 
\tilde{P}_{N+L_2 - M_2}(\mu_k)  & \dots & \tilde{P}_{N+L_1 - M_1-1}(\mu_k) }
\label{Gmatrix}
\ee
with
\bea
\ds{\mathop{
K_{12}}^{\!J}}(\xi, \zeta) &\&:= \sum_{n=0}^{J-1} P_n(\xi) S_n(\zeta) 
\label{K12}\\
\ds{\mathop{
K_{11}}^{\!J}}(\xi, \eta) &\&:= \sum_{n=0}^{J-1} P_n(\xi) {\tilde S_n}(\eta)
+ {1\over \xi - \eta}
\label{K11}\\
\ds{\mathop{
K_{22}}^{\!J}}(\mu, \zeta) &\&:= \sum_{n=0}^{J-1} {\tilde P}_n(\mu) S_n(\zeta)
+ {1\over \zeta - \mu }
\label{K22}\\
\ds{\mathop{
K_{21}}^{\!J}}(\mu, \eta) &\&:= \sum_{n=0}^{J-1} {\tilde P}_n(\mu)  {\tilde S}_n(\eta)
- H(\mu, \eta) 
\label{K21}\\
H(\mu, \eta) &\&:=  \int {d\mu(x,y) \over (\eta-x)(\mu-y)}.
\label{HH}
\eea
Similar expressions for the cases when one or both of the inequalities (\ref{LNMinequal}) is reversed will also be derived. (See Section 3.)

   Within the setting  of two-matrix models,  the integral (\ref{mainintegral}) may be interpreted as the  expectation value  of the product of $L_1$ evaluations of the characteristic polynomial of the first matrix and $L_2$ of that of the second matrix, divided by similar products of $M_1$ and $M_2$ further evaluations of the respective characteristic polynomials. Integrals of this type were computed in the context of complex 
 matrix models, where the variables  $(x_a,y_a)$ are replaced by pairs $(z_a,\bar{z}_a)$ of complex conjugate values,  in \cite{AV}  for the special case when $M_1=M_2=0$,  and, more generally, in \cite{Be}, for all values of $(L_1,L_2,M_1, M_2)$. The method of derivation used by these authors was based essentially upon recursive arguments, and is rather lengthy compared with the ``direct method'' that we present here. Other cases of the integral  (\ref{mainintegral}), in which the condition (\ref{LNMinequal}) does not hold, give rise to analogous determinantal expressions. These too will be derived in Section 3, using the same methods as those leading to (\ref{mainintegral}).
 
 \br
 Note that the multiplicative factor in front of the $\det g$ term in (\ref{mainintegral}) is just a product of  the inverse determinants that enter into rational interpolation formulae;
 e.g., if $L_1\ge M_1$, $L_2 \ge M_2$, then 
 \be
{ \prod_{\alpha=1}^{L_1}\prod_{j=1}^{M_1}(\xi_\alpha - \eta_j)\prod_{\beta=1}^{L_2}\prod_{k=1}^{M_2}(\zeta_\beta- \mu_k)
 \over \Delta_{L_1}(\xi)  \Delta_{L_2}(\zeta)  \Delta_{M_1}(\eta)  \Delta_{M_2}(\mu)}
 ={1\over \det G_1  \det G_2}
 \ee
 where
 \be 
 G_1 := \pmatrix {{1\over \xi_\alpha - \eta_j} & \xi_\alpha^b}_{1\le \alpha \le L_1,\ 
  1\le j \le M_1 \atop
 b\le L_1-M_1 -1},
 \quad
  G_2 := \pmatrix {{1\over \zeta_\beta - \mu_k} & \zeta_\beta^c}_{1\le \beta \le L_2,\ 
  1\le j \le M_2 \atop
 c\le L_2-M_2 -1},
  \ee
  
 \er
\br
Expressions for the biorthogonal polynomials $\{P^{(\xi,\zeta,\eta,\mu)}_n(x), S^{(\xi,\zeta,\eta,\mu)}_n(y)\}_{n\in \Nb}$ with respect to the modified measure
\be
d\mu_{\xi,\zeta,\eta.\mu}(x,y)
 \prod_{a=1}^N {\prod_{\alpha=1}^{L_1}(\xi_\alpha - x_a) \prod_{\beta=1}^{L_2}(\zeta_\alpha - y_a) \over \prod_{j=1}^{M_1}(\eta_j - x_a)  \prod_{k=1}^{M_2}(\mu_k - y_a)} d\mu(x,y),
 \ee
 together with their Hilbert transforms  $\{\tilde{P}^{(\xi,\zeta,\eta,\mu)}_n(x), \tilde{S}^{(\xi,\zeta,\eta,\mu)}_n(y)\}_{n\in \Nb}$ may immediately be  deduced
 from formula (\ref{mainintegral}),  simply by making the respective replacements
 \bea
 \prod_{\alpha=1}^{L_1}(\xi_\alpha - x_a) &\& \ra (x-x_a) \prod_{\alpha=1}^{L_1}(\xi_\alpha - x_a), \qquad
 \prod_{\beta=1}^{L_2}(\zeta_\beta - y_a)
 \ra (y-y_a) \prod_{\beta=1}^{L_2}(\zeta_\beta - y_a), \cr
\prod_{j=1}^{M_1}(\eta_j - x_a) &\& \ra (x-x_a)\prod_{j=1}^{M_1}(\eta_j - x_a),  \qquad
\prod_{k=1}^{M_2}(\mu_k - x_a) \ra (y-y_a) \prod_{k=1}^{M_2}(\mu_k - x_a)
 \eea
in  $\Ib^{(2)}_n$ and multiplying the result by the normalizing factor
$\sqrt{\Ib^{(2)}_{n+1} \over  \ (n + 1) \Ib^{(2)}_n}$. 
\er
\br
The corresponding formulae for the cases where one or more of the parameters within the groups $\{\xi_\alpha\}$, $\{\zeta_\beta\}$, $\{\eta_j\}$,  $\{\mu_k\}$ coincide may be very easily determined from (\ref{mainintegral}), simply by taking the appropriate limits.
This replaces the terms which appear multiply in the entries in (\ref{Gmatrix}) and  the sums (\ref{K12}) - (\ref{K21}) by their derivatives with respect to the repeated parameters.
\er
   
   As a ``warm-up exercise'', we also rederive the simpler analogous results for integrals of symmetric rational functions in one set of $N$ variables arising, e.g., in  Hermitian one-matrix models. In this case, let $\{P_n(x)\}_{n=0,1 \dots}$ denote the orthogonal polynomials with respect to a measure $d \mu(x)$, which may again, in general, be complex and supported on an arbitrary finite union of curve segments:
   \be
   \int d \mu(x) P_n(x) P_m(x) = \delta_{nm},
   \label{orthogonality}
   \ee
  with leading term normalization
   \be
    P_n(x) = {x^n\over\sqrt{h_n}} + O(x^{n-1}).
\ee
To guarantee their existence, the finiteness of the Hankel matrix of moments
\be
M_{jk} := \int d\mu(x)x^ {j+k}<\infty, \quad 0 , \quad \forall j,k \in {\bf N}.
\ee
is assumed, as well as the nonsingularity of all diagonal $N \times N$ submatrices $(M_{jk})_{0\le j, k \le N-1}$. 
The Hilbert transforms 
\be
\tilde{P}_n(\eta):= \int d\mu(x) {P_n(x) \over \eta -x}
\ee
 of the orthogonal polynomials are again assumed to exist.
  The partition function is
\be
{\bf Z}_N := \int d(\mu)x_1) \dots \int d\mu(x_N) \Delta_N^2(x) = N!\prod_{n=0}^{N-1}h_n.
\label{1matrixpartfn}
\ee
 The  expression analogous to (\ref{mainintegral}) for the integral of a symmetric, rational function in the  $N$ variables $\{x_a\}_{a=1,\dots N}$, having simple zeros at the points  $\{\xi_\alpha\}_{\alpha = 1 \dots L}$,  and simple poles at $\{\eta_j\}_{j= 1 \dots M}$,
 valid for $N \ge M $, is
     \bea
 {\bf  I}_N &\&:= {1\over {\bf Z}_N} \int d\mu(x_1) \dots \int d\mu(x_N) \prod_{a=1}^N {\prod_{\alpha=1}^L (\xi_\alpha - x_a)\over \prod_{j=1}^M (\eta_j - x_a)}
   \Delta_N^2(x)  
   \label {singleintegral_def} \\
   &\&= 
 { (-1)^{{M(M-1)\over 2}+LM} \prod_{n=N}^{N+L-1}\sqrt{h_n}\over \Delta_L(\xi) \Delta_M(\eta) 
  \prod_{n=N-M}^{N - 1}\sqrt{h_n}} 
           \det\pmatrix{P_{N-M}(\xi_\alpha) & \dots & P_{N+L-1}(\xi_\alpha) \cr 
        \tilde{P}_{N-M}(\eta_j) & \dots & \tilde{P}_{N+L-1}(\eta_j)}, 
   \label{singleintegralNgM}
   \eea
    where the two groups of complex parameters $\{\xi_\alpha\}_{\alpha=1, \dots L}$,
   $\{\eta_j\}_{j=1, \dots M}$ are assumed to have distinct values, the latter being evaluated off the contours of integration.
   The analogous formula for the same integral, valid for $N < M$ is
       \bea
 {\bf  I}_N &\& =
{(-1)^{{1\over 2}N(N-1)}(-1)^{LM}\prod_{n=0}^{M-N}\sqrt{h_n} \prod_{n=N}^{N+L-1}\sqrt{h_n}
\over \Delta_L(\xi) \Delta_M(\eta) \prod_{n=0}^{N-1} \sqrt{h_n}} \times
\cr
&\& \quad \times
\det\pmatrix{P_0(\xi_\alpha) & \dots & P_{N+L-1}(\xi_\alpha) & 0 & \dots & 0 
\cr
\tilde{P}_0(\eta_j) & \dots & \tilde{P}_{N+L-1}(\eta_j) & P_0(\eta_j) & \dots  &
 P_{M-N}(\eta_j) }.  
  \label{singleintegralMgN}
   \eea
     
   Such relations for polynomial integrands originated in the work of Heine and Christoffel \cite{G, Sz} and were extended to the general rational case in \cite{U}. The case
(\ref{singleintegralNgM}) was recently rederived  in the context of Hermitian matrix models by other methods \cite{BDS, BH, FS}.  The direct method given here leads to relations  (\ref{singleintegralNgM}), (\ref{singleintegralMgN}) in just a few lines.

The key tool that is used in our ``direct'' approach is the following identity,  which is just a multivariable partial fraction expansion for rational symmetric functions, 
valid if $N\ge M$:
  \be
 {\Delta_N(x) \Delta_M(\eta) \over \prod_{a=1}^N \prod_{j=1}^M(\eta_j -x_a)}
  = (-1)^{MN} \sum_{\sigma\in S_M}\sgn(\sigma)\!\!
  \sum_{a_1<  \dots < a_M}^N \!\!(-1)^{\sum_{j=1}^M a_j}
{  \Delta_{N-M}(x[{\bf a}])\over  \prod_{j=1}^M(\eta_{\sigma_j}- x_{a_j})}.
\label{partialfrac1}
  \ee
  Here $x[{\bf a}]$ denotes the sequence  $(x_1, \dots x_N)$ with the elements
  $(x_{a_1}, \dots x_{a_M})$ omitted, and $\sgn(\sigma)$ is the sign of the permutation
 $\sigma = \pmatrix{ 1 & \dots & M \cr \sigma_1 & \dots  & \sigma_M} \in S_M$. (To verify this identity, one simply notes that, viewed as rational functions in the $\eta_j$'s,  the residues at all poles coincide, and both sides tend to $0$ as $\eta_j \ra \infty$
  for any $j$.) Reversing the r\^oles of $\{x_a\}$ and $\{\eta_j\}$,  an equivalent identity  of slightly different form is valid, in this case for  $N \le M$:
  \be
   {\Delta_N(x) \Delta_M(\eta) \over \prod_{a=1}^N \prod_{j=1}^M(\eta_j -x_a)}
  = {(-1)^{{1\over 2}N(N-1)}\over (M-N)!} \sum_{\sigma\in S_M}\sgn(\sigma)
{\Delta_{M-N}(\eta_{\sigma_{N+1}}, \dots, \eta_{\sigma_{M}})\over  \prod_{a=1}^N(\eta_{\sigma_a}- x_a)},
\label{partialfrac2}
  \ee
  
  Another relation that is of importance is the Cauchy-Binet identity from multilinear algebra. In invariant form, this states that if $V$ is an oriented  Euclidean vector space with volume form $\Omega$, and we have two sets of $L$ vectors
   $(P^1, \dots ,P^L)$, $(S^1, \dots ,S^L)$, then the scalar product of their exterior products $(\wedge_{\alpha=1}^{L}P^\alpha,\wedge_{\beta=1}^{L}S^\beta)$, defined by
   \be
     \wedge_{\alpha=1}^{L}P^\alpha *\wedge_{\beta=1}^{L}S^\beta=
  (\wedge_{\alpha=1}^{L}P^\alpha,\wedge_{\beta=1}^{L}S^\beta)\Omega
      \label{CauchyBinetwedge}
     \ee
     (where $*$ denotes Hodge dual on exterior forms)
     is equal to the determinant of the matrix formed from the scalar products
     \bea
       (\wedge_{\alpha=1}^{L}P^\alpha ,\wedge_{\beta=1}^{L}S^\beta) &\&= \det G  \\
       G^{\alpha \beta} &\&:= (P^\alpha, S^\beta), \quad 1\le i, j \le L
       \eea
In component form, if $\dim V = N+L$, and the vectors  
$\{P^\alpha, S^\alpha\}_{1\le \alpha \le L}$ have components
$\{P^\alpha_j, S^\alpha_k\}_{1\le j,k \le L+N}$ 
relative to a positively oriented orthonormal frame, this reads:
       \be
       \epsilon^{j_1 \dots j_N j_{N+1} \dots j_{N+L}} \epsilon^{j_1 \dots j_N k_{N+1} \dots k_{N+L}}
       P^1_{j_{N+1}} \dots P^L_{j_{N+L}} S^1_{k_{N+1}} \dots S^L_{k_{N+L}} = N! \det(G),
       \label{CauchyBinetcomp}
            \ee
 where $\epsilon$ denotes the Levi-Civita symbol and paired indices are again summed over.

It is worthwhile noting  that both ({\ref{partialfrac1}) and (\ref{CauchyBinetwedge}) are particular forms of determinantal identities that may be deduced from Wick's theorem  for products of free Fermi field operators.  In a sequel to this work  \cite{HO2}, another method of deriving the relations (\ref{mainintegral}), ({\ref{singleintegralNgM}) and ({\ref{singleintegralMgN}) is given, based directly on evaluation of vacuum state matrix elements of operators constructed from  products and exponentials of Fermionic free fields.
  
\section{The one matrix case. Proof of eqs.  (\ref{singleintegralNgM}), (\ref{singleintegralMgN})}

   To introduce the ``direct'' method used here, we begin by deriving
 the relation  (\ref{singleintegralNgM}). First, we recall the proof for the case $M=0$ given in \cite{BH}. 
    \bea
    {\bf  I}_N  &\& =  {1\over {\bf Z}_N} \int d \mu(x_1) \dots \int d \mu(x_N) \prod_{a=1}^N \prod_{\alpha=1}^L(\xi_\alpha - x_a)\Delta^2_N(x) \cr
&\&=   {1\over N! (\prod_{n=0}^{N-1}h_n)  \Delta_L(\xi)} \int d \mu(x_1) \dots \int d \mu(x_N)\Delta_{N+L}(x,\xi) \Delta_N(x) \cr
&\&=  { \prod_{n=N}^{N+L-1}\sqrt{h_n}\over N! \Delta_L(\xi)}
\int d \mu(x_1) \dots \int d \mu(x_N) \det\pmatrix{P_j(x_a) & P_j(\xi_\alpha)}
\det(P_k(x_b))\cr
&\&= { \prod_{n=N}^{N+L-1}\sqrt{h_n}\over N! \Delta_L(\xi)}\int d \mu(x_1) \dots \int d \mu(x_N) \epsilon^{j_1 \dots j_N j_{N+1} \dots j_{N+L}}
P_{j_1}(x_1) \dots P_{j_N}(x_N) \cr
&\& \qquad \times
P_{j_{N+1}}(\xi_1)  \dots P_{j_{N+L}}(\xi_L)\epsilon^{k_1 \dots k_N} P_{k_1}(x_1) \dots P_{k_N}(x_N).
\eea
Here the summation convention is used and 
$ \epsilon^{j_1 \dots j_N j_{N+1} \dots j_{N+L}}$ and $\epsilon^{k_1 \dots k_N}$
denote the Levi-Civita symbol  in $N+L$ and $N$ variables, respectively,
the ranges of summation being $0\le j, j_1, \dots , j_{N+L} \le  N+L-1$ and $0 \le k, k_1, \dots ,k_L \le  N-1$.
Using the orthogonality relations (\ref{orthogonality}) to evaluate the integrals yields
\be
{\bf I}_N  =  {\prod_{n=N}^{N+L-1} \sqrt{h_n} \over  \Delta_L(\xi)}
\det( P_{N+\alpha-1}(\xi_\beta))_{1\le \alpha, \beta \le L}.
\label{numeratorcase}
\ee

  To extend this result to the case of arbitrary $M$, we make use of the
  identity (\ref{partialfrac1}). Substituting this into the integrand of (\ref{singleintegralNgM}),  using  symmetry under permutations of the $x_a$'s, and 
   invariance under relabeling of the integration variables gives
     \bea
    {\bf  I}_N &\&= {(-1)^{M(M-1)\over 2} N! \over { \bf Z}_N \Delta_M(\eta)(N-M)! }\int {d\mu(z_1) \over \eta_1 - z_1} \dots\int {d\mu(z_M) \over \eta_j - z_j}
   \Delta_M(z)  \prod_{\alpha =1}^L\prod_{j=1}^M (\xi_\alpha - z_j) \cr
    &\&\quad \times  \int d\mu(x_1) \dots \int d\mu(x_{N-M})  \Delta^2_{N-M}(x_1, \dots ,x_{N-M})   \prod_{a=1}^{N -M}  \prod_{\alpha=1}^L(\xi_\alpha - x_a)\prod_{j=1}^M (z_j - x_a),
         \cr
     &\&
     \label{oneintegral_int}
      \eea 
where the combinatorial factor ${ N! \over(N-M)!}$ has been introduced to replace the sum over all permutations of ordered choices of $M$ of the elements $\{x_a\}_{a=1, \dots, N}$, which are here relabeled $(z_1, \dots , z_M)$, while the remaining
      ones are relabelled $(x_1, \dots , x_{N-M})$.
  Now, applying the relation (\ref{numeratorcase}) with the $L$ parameters
  $\xi :=(\xi_1, \dots ,\xi_L)$  extended to $L+M$ parameters $(\xi, z):=(\xi_1, \dots ,\xi_L, z_1, \dots ,z_M)$ to (\ref{oneintegral_int}), and $N$ replaced by $N-M$, and using (\ref{1matrixpartfn}) for both ${\bf Z}_N$ and ${\bf Z}_{N-M} $ gives
  \bea
  {\bf I}_N &\&=  {(-1)^{{M(M-1)\over 2}} (-1)^{LM} \prod_{m=N-M}^{N+L-1}\sqrt{h_m}\over \Delta_L(\xi) \Delta_M(\eta) \prod_{n=N-M}^{N-1}h_n }\int {d\mu(z_1) \over \eta_1 - z_1} \dots\int {d\mu(z_M) \over \eta_j - z_j}\cr
  &\& \quad \times
  \det\pmatrix{P_{N-M}(\xi_\alpha) & \dots & P_{N+L-1}(\xi_\alpha) \cr 
        P_{N-M}(z_j) & \dots & P_{N+L -1}(z_j)}.
\eea
The multilinearity of the determinant when evaluating the integrals
then gives  ({\ref{singleintegralNgM}).

Now consider the case $N<M$. Substituting (\ref{partialfrac2})  into the definition 
(\ref{singleintegral_def}) of ${\bf I}_N$ and using (\ref{1matrixpartfn}) gives
\bea
{\bf I}_N &\& = {(-1)^{{1\over 2}N(N-1)}\over N! (M-N)! \Delta_L(\xi) \Delta_M(\eta)}
\cr
&\& \quad \times
\sum_{\sigma\in S_M} \int d\mu(x_1) \dots \int d\mu(x_N) 
{\Delta_{N+L}(x, \xi)\Delta_{M-N}(\eta_{\sigma_{N+1}},\dots, \eta_{\sigma_M})
\over \prod_{a=1}^N(\eta_{\sigma_a} - x_a)}\cr
&\& = {(-1)^{{1\over 2}N(N-1)}\prod_{n=0}^{M-N}\sqrt{h_n} \prod_{n=N}^{N+L-1}\sqrt{h_n}
\over N! (M-N)!  \prod_{n=0}^{N-1} \sqrt{h_n}\Delta_L(\xi) \Delta_M(\eta)}
\det\left({P_j(\eta_{\sigma_{N+k}})_{0 \le j \le M-N-1 \atop 1\le k \le M-N}}\right)
\cr
&\& \quad \times
\sum_{\sigma\in S_M} \int {d\mu(x_1)\over (\eta_{\sigma_1} - x_1)} \dots \int {d\mu(x_N)  \over (\eta_{\sigma_N} - x_N)}
\det\pmatrix{P_l(x_a)\vert_{0 \le l \le N+L-1 \atop 1\le a \le N}\cr
P_l(\xi_\alpha)\vert_{0 \le l \le N+L-1 \atop 1\le \alpha \le L}} \cr 
&\& = {(-1)^{{1\over 2}N(N-1)}\prod_{n=0}^{M-N}\sqrt{h_n} \prod_{n=N}^{N+L-1}\sqrt{h_n}
\over N! (M-N)!  \prod_{n=0}^{N-1} \sqrt{h_n}\Delta_L(\xi) \Delta_M(\eta)}
\cr
&\& \quad \times \sum_{\sigma\in S_M}
\det\left({P_j(\eta_{\sigma_{N+k}})\vert_{0 \le j \le M-N-1 \atop 1\le k \le M-N}}\right)
 \det\pmatrix{\tilde{P}_l(\eta_{\sigma_a})\vert_{0 \le l \le N+L -1\atop 1\le a \le N}\cr
P_l(\xi_\alpha)\vert_{0 \le l \le N+L -1 \atop 1\le \alpha \le L}},
\eea
which is just the block matrix determinantal expansion of 
 ({\ref{singleintegralNgM}).

\section{The coupled matrix case}
\subsection {Case 1. $N + L_1-M_1 \ge N +L_2 -M_2 \ge 0$ }

   We now turn to the derivation of relation (\ref{mainintegral}), which will be
   done along similar lines to the above. We  again begin with the case with
   no denominator factors; i.e. $M_1=M_2 =0$, and $L_1\ge L_2$. For this case,
   using (\ref{partfn}),
   \bea
  {\bf I}^{(2)}_N &\&:= {1\over {\bf Z}^{(2)}_N} \prod_{a=1}^N\left(\int    d\mu(x_a,y_a) 
\prod_{\alpha=1}^{L_1}(\xi_\alpha - x_a) \prod_{\beta=1}^{L_2}(\zeta_\alpha - y_a)\right)
 \Delta_N(x)\Delta_N(y)  \cr
 &\& =  {1\over N! (\prod_{n=0}^{N-1}h_n)\Delta_{L_1}(\xi) \Delta_{L_2}(\zeta)}
 \prod_{a=1}^N \left( \int   d\mu(x_a,y_a)\right)
 \Delta_{N+L_1}(x, \xi)\Delta_{N+L_2}(y, \zeta) \cr
 &\& =
   {\prod_{n=N}^{{}_{N+L_2-1}}\sqrt{h_n} \prod_{n=N}^{{}_{N+L_1-1}}\sqrt{h}_n\over N!\Delta_{L_1}(\xi) \Delta_{L_2}(\zeta)} \cr
   &\& \quad \times
   \prod_{a=1}^N \left( \int  d\mu(x_a,y_a)\right)
 \det(P_j(x_b) \ P_j(\xi_\alpha)) \det(S_k (y_c) \  S_k(\zeta_\beta))
  \cr
  &\& =
   {\prod_{{}_{n=N}}^{{}_{N+L_2-1}}\sqrt{h_n} \prod_{{}_{n=N}}^{{}_{N+L_1-1}}\sqrt{h}_n\over N!\Delta_{L_1}(\xi) \Delta_{L_2}(\zeta)} \cr
   &\& \quad \times \epsilon^{j_1 \dots j_N j_{N+1} \dots j_{N+L_1}}
    P_{j_1}(x_1)  \dots P_{j_N}(x_N)
  P_{j_{N+1}}(\xi_1) \dots P_{j_{N+L_1}}(\xi_{L_1})\cr
   &\& \quad \times \epsilon^{j_1 \dots j_N k_{N+1} \dots k_{N+L_2}}
    S_{j_1}(y_1)  \dots S_{j_N}(y_N)
  S_{k_{N+1}}(\zeta_1) \dots S_{k_{N+L_2}}(\zeta_{L_2}),
  \label{epsilonform}
   \eea
where in the first determinant $0\le j, j_1, \dots , j_{N+L_1} \le N+L_1-1$,  and in the second $0\le k, k_{N+1}, \dots ,k_ {N+L_2}\le  N+L_2-1 $. To complete the computation, we now make use of the Cauchy-Binet identity (\ref{CauchyBinetcomp}). 
 To apply this to the expression (\ref{epsilonform}), let $L:=L_1$ and identify the vectors $\{P^\alpha, S^\beta\}_{1 \le \alpha, \beta \le L}$
       as follows:
       \bea
       P^\alpha_j &\&:= P_{j-1}(\xi_\alpha), \qquad 1\le i \le L_1,  \qquad 0\le j \le N+L_1-1, \cr
        S^\beta_j &\&:= S_{j-1}(\zeta_\beta), \qquad 1\le \beta \le L_2, \qquad 0\le j \le N+ L_2-1, \cr
           S^\beta_j &\&:= 0 , \qquad\qquad \quad  1\le \beta \le L_2, \quad N+L_2\le j \le N+ L_1-1, \cr
         S^\beta_j &\&:= \delta_{N+L_2 +\beta-1, j}, \quad L_2+1\le \beta \le L_1, \quad 0\le j \le N+L_1-1.
         \eea
      Using the Cauchy-Binet identity (\ref{CauchyBinetcomp}) and eq.~(\ref{partfn}), the equality
   (\ref{epsilonform}) gives the following expression for  ${\bf I}^{(2)}_N$.
       \be
         {\bf I}^{(2)}_N = 
          {\prod_{n=N}^{{}_{N+L_2-1}}\sqrt{h_n} \prod_{n=N}^{{}_{N+L_1-1}}\sqrt{h}_n \over \Delta_{L_1}(\xi) \Delta_{L_2}(\zeta)}
          \det\pmatrix{\ds{\mathop{
K_{12}}^{\!N+L_2}} (\xi_\alpha, \zeta_\beta) & P_{N+L_2}(\xi_\alpha) \dots P_{N+L_1-1}(\xi_\alpha)},
\label{mainintegralzeroM}
\ee  
where
\be
\mathop{
K_{12}}^{\!\!N+L_2}(\xi, \zeta) := \sum_{n=0}^{N +L_2-1}P_n(\xi) S_n(\zeta).
\ee
   
   We now extend this result to the case of arbitrary $(L_1,L_2,M_1,M_2)$  satisfying
 (\ref{LNMinequal}).   We detail the derivation only in the case when the stronger inequality
   \be
   N  \ge \max(M_1, \ M_2)
\label{NMinequal}
\ee
holds. For the intermediate cases, when $N$ lies between $M_1-L_1$ and
$M_1$ or between $M_2-L_2$ and $M_2$,  formula (\ref{mainintegral}) may be derived by similar computations. 
   
   Substituting the identity (\ref{partialfrac1}) for both 
   denominator factors $\prod_{a=1}^N\prod_{j=1}^{M_1}(\eta_j - x_a)$
   and $\prod_{a=1}^N\prod_{k=1}^{M_2}(\mu_k - y_a)$ into the integral (\ref{I2Ndef})
   defining ${\bf I}_N^{(2)}$ gives:
   \bea
   {\bf I}^{(2)}&\&=   {(-1)^{(M_1+M_2)N} \over {\bf Z}^{(2)}_N \Delta_{M_1}(\eta)\Delta_{M_2}(\mu)}\sum_{\sigma\in S_{M_1}}{\hskip -5 pt}\sgn(\sigma)
   \sum_{\tilde{\sigma}\in S_{M_2}}{\hskip -5 pt}\sgn(\tilde{\sigma})\sum_{a_1<  \dots < a_{M_1}}^N{\hskip -10 pt} (-1)^{\sum_{j=1}^{M_1} a_j}
\sum_{b_1<  \dots < b_{M_2}}^N {\hskip -10 pt}(-1)^{\sum_{k=1}^{M_2} b_k}
  \cr
 &\& \quad  \times \prod_{a=1}^N\left( \int  d\mu(x_a,y_a) 
  \prod_{\alpha=1}^{L_1}(\xi_\alpha - x_a) \prod_{\beta=1}^{L_2}(\zeta_\beta - y_a)\right)
 {\Delta_{N-M_1}(x[{\bf a}])\Delta_{N-M_2}(y[{\bf b}])
  \over \prod_{j=1}^{M_1}(\eta_{\sigma_j}- x_{a_j})  
\prod_{k=1}^{M_2}(\mu_{\tilde{\sigma}_k}- y_{b_k})  } .\cr
  &\&
  \label{I2_intermediate}
 \eea
 In this sum we must distinguish:
 \begin{list}{}
 \item{} $m:=$ the number of $a_j$'s that coincide with $b_k$'s
 \item{} $M_1-m =$ the number of $a_j$'s that do not coincide with any $b_k$'s
  \item{} $M_2-m =$ the number of $b_k$'s that do not coincide with any $a_j$'s
  \end{list}
\noindent 
Note that, if $M_1 + M_2 \le N$, $m$ can vary from $0$ to $\min(M_1, M_2)$, but
if $N < M_1 + M_2$, it can only take values $ m\ge M_1 + M_2 - N$.
The number of distinct ways in which two such ordered sets $a_1<\dots < a_{M_1}$,   and \hbox{$b_1<\dots < b_{M_1}$} with exactly $m$ common elements can be chosen from
the numbers $(1, \dots ,N)$ is: 
  \be
 C^N_{m,M_1,M_2}:= {N!\over (N-M_1 - M_2+m)!(M_1-m)!(M_2-m)! m!}.
 \label{combconst}
  \ee
  In view of the invariance of the integrand in (\ref{mainintegral}) under permutations
  of the pairs $\{(x_a,y_a)\}_{a=1, \dots N}$, and the freedom to relabel the integration variables, we may express the integral as
 \bea
   {\bf I}^{(2)}&\&=   { (-1)^{{1\over 2}(M_1+M_2)(M_1+M_2 -1)}\over {\bf Z}^{(2)}_N \Delta_{M_1}(\eta)\Delta_{M_2}(\mu)}
 \sum_{m=\max(0, M_1+M_2-N)}^{\min(M_1,M_2)}   (-1)^{m}
   \sum_{\sigma\in S_{M_1}}{\hskip -5 pt}\sgn(\sigma)
   \sum_{\tilde{\sigma}\in S_{M_2}}{\hskip -5 pt}\sgn(\tilde{\sigma})  \cr
 &\& \quad  \times
   C^N_{m,M_1,M_2}
 \prod_{i=1}^{m}\left( \int d\mu(z_i,w_i) 
  {\prod_{\alpha=1}^{L_1}(\xi_ \alpha - z_i) \prod_{\beta=1}^{L_2}(\zeta_\beta - w_i) \over (\eta_{\sigma_i}  - z_i ) 
(\mu_{\tilde{\sigma}_i}  - w_i)  }\right)\cr
&\& \quad  \times 
  \prod_{j=1}^{M_1-m} \left( \int  d\mu(\tilde{z}_j, \tilde{w}_j)  
  {\prod_{\alpha=1}^{L_1}(\xi_\alpha - \tilde{z}_j) \prod_{\beta=1}^{L_2}(\zeta_\beta - \tilde{w}_j) \over \eta_{\sigma_{m+j}}  - \tilde{z}_j  }
  \right) \Delta_{M_1-m}(\tilde{w})\cr
&\& \quad  \times 
 \prod_{k=1}^{M_2-m}\left( \int   d\mu(\tilde{\tilde{z}}_k,\tilde{\tilde{w}}_k) 
  {\prod_{\alpha=1}^{L_1}(\xi_\alpha - \tilde{\tilde{z}}_k) \prod_{\beta=1}^{L_2}(\zeta_\beta - \tilde{\tilde{w}}_k) \over  \mu_{\tilde{\sigma}_{M_1+k}} -  \tilde{\tilde{w}}_k  }\right)
  \Delta_{M_2-m}(\tilde{\tilde{z}})\cr
&\& \quad  \times
  {\hskip -10 pt} \prod_{a=1}^{{}_{\!N-\!M_1-\!M_2+m}}\left(  \int d\mu(x_a,y_a) 
  \prod_{\alpha=1}^{L_1}(\xi_\alpha - x_a) \prod_{k=1}^{M_2-m}(\tilde{\tilde{z}}_k- x_a)
   \prod_{\beta=1}^{L_2}(\zeta_\beta  - y_a) \prod_{j=1}^{M_1-m}(\tilde{w}_j  - y_a)
   \right),  \cr
 &\&  \quad \times {\hskip 60 pt} \Delta_{{}_{\!N-\!M_1-\!M_2+m}}(x)\Delta_{{}_{\!N-\!M_1-\!M_2+m}}(y)
     \label{fullsigmasum}
     \eea
where we have made the following changes of notation in the integration variables
\bea
\pmatrix{x_1, \dots , x_m \cr
             y_1, \dots , y_m}
              &\& \ra  \pmatrix{z_1, \dots , z_m \cr
             w_1, \dots , w_m}  \cr
             \pmatrix{x_{m+1}, \dots , x_{M_1} \cr
             y_{m+1}, \dots , y_{M_1}}
              &\& \ra  \pmatrix{\tilde{z}_1, \dots , \tilde{z}_{M_1-m} \cr
             \tilde{w}_1, \dots , \tilde{w}_{M_1-m} }  \cr
             \pmatrix{x_{M_1+1}, \dots , x_{M_1+M_2-m} \cr
             y_{M_1 +1}, \dots ,y_{M_1+M_2-m} }
              &\& \ra  \pmatrix{\tilde{\tilde{z}}_1, \dots , \tilde{\tilde{z}}_{M_2-m} \cr
            \tilde{ \tilde{w}}_1, \dots ,\tilde{ \tilde{w}}_{M_2-m} }  \cr
\pmatrix{x_{M_1+M_2-m+1}, \dots , x_N \cr
             y_{M_1+M_2-m+1} \dots , y_N}
              &\& \ra  \pmatrix{x_1 , \dots , x_{N-M_1-M_2 +m} \cr
             y_1 \dots , y_{N-M_1-M_2 +m}}. 
  \eea
 In determining the sign factor in the first line of (\ref{fullsigmasum}), we have replaced the  sums 
   $\sum_{j=1}^{M_1} a_j$ and $\sum_{k=1}^{M_2} b_j$ 
   appearing in (\ref{I2_intermediate}) by their values for the case 
   $(a_1, \dots , a_{M_1}) = (1, \dots , M_1)$  and  $(b_1, \dots , b_{M_2}) 
   = (1, \dots , m, M_1+1, \dots , M_1 + M_2)$. 
   This leaves a residual factor $(-1)^{m^2}=(-1)^m$, giving an alternating sign in the sum over $m$.  
   
  A further simplification can be made in (\ref{fullsigmasum}) by noting 
  that, in the sums over the elements of the symmetric groups $S_{M_1}$
  and $S_{M_2}$, all terms in the integrations over the $\{z_i,w_i\}_{i=1 \dots m}$,
  $\{\tilde{z}_j,\tilde{w}_j\}_{j=1 \dots M_1-m}$ and $\{\tilde{\tilde{z}}_j,\tilde{\tilde{w}}_k\}_{k=1 \dots M_2-m}$
  variables coming from pairs of permutations $(\sigma, \tilde{\sigma})$ for
  which the sets $\{(\sigma_i, \tilde{\sigma}_i)\}_{i=1, \dots m}$,
  $\{\sigma_{m+j}\}_{k=1, \dots M_1-m}$ and   $\{\tilde{\sigma}_{M_1+k}\}_{k=1, \dots M_2-m}$  are invariant  contribute the same value to the sum, and there are 
  \hbox{$m!(M_1-m)!(M_2-m)!$} of these.  That is, all left cosets of the subgroup 
  $S_{m, M_1,M_2} :=  S_m \times S_{M_1-m} \times S_{M_2-m} \subset S_{M_1} \times S_{M_2}$ that permute separately the first $m$ elements in both $(1, \dots , M_1)$ and $(1, \dots , M_2)$, the last $M_1-m$ elements in $(1, \dots , M_1)$ and  the last $M_2-m$ elements in $(1, \dots , M_2)$ contribute the same
  term in the sum. Hence we may choose one representative 
   $[\sigma, \tilde{\sigma}] \in (S_{M_1} \times S_{M_2})/S_{m, M_1,M_2}$
 from each coset, multiplying this term by the factor $m!(M_1-m)!(M_2-m)!$.
 
    We may also now apply relation (\ref{mainintegralzeroM}), with the replacements 
 $N\ra N -M_1-M_2+m$, \hbox{$L_1\ra L_1 + M_2-m$}, $L_2 \ra L_2+M_1-m$,
 as well as the expression (\ref{partfn}) for both ${\bf Z}_N^{(2)}$ and
  ${\bf Z}_{N- M_1 - M_2 +m}^{(2)}$ to
 evaluate the integrals over the  $\{(x_a, y_a)\}_{a=1 \dots N-M_1-M_2 +m}$ variables. Using (\ref{partfn}), the resulting sum becomes
  \bea
   {\bf I}^{(2)}&\&=   {(-1)^{{1\over 2}(M_1+M_2)(M_1+M_2 - 1)} \prod_{n=N}^{N+L_2-M_2-1}\sqrt{h_n}
   \prod_{n=N}^{N+L_1 -M_1-1}\sqrt{h_n} 
\over N!  \Delta_{L_1}(\xi)\Delta_{L_2}(\zeta)\Delta_{M_1}(\eta)\Delta_{M_2}(\mu)}  \cr
&\&  \qquad \times  \sum_{m=0}^{\min(M_1,M_2)}  (-1)^m 
 \sum_{[\sigma,\tilde{\sigma}]   }\sgn(\sigma) \sgn(\tilde{\sigma})
\prod_{i=1}^{m}\left( \int  d\mu(z_i,w_i) 
  {\prod_{\alpha=1}^{L_1}(\xi_ \alpha - z_i) \prod_{\beta=1}^{L_2}(\zeta_\beta - w_i) \over (\eta_{\sigma_i}  - z_i ) 
(\mu_{\tilde{\sigma}_i}  - w_i)  }\right)\cr
&\& \quad  \times 
  \prod_{j=1}^{M_1-m}\left( \int  d\mu(\tilde{z}_j, \tilde{w}_j)  
  {\prod_{\alpha=1}^{L_1}(\xi_\alpha \!- \!\tilde{z}_j) \over \eta_{\sigma_{m+j}}  - \tilde{z}_j  }\right)
  \prod_{k=1}^{M_2-m}\left( \int  d\mu(\tilde{\tilde{z}}_k,\tilde{\tilde{w}}_k) 
  { \prod_{\beta=1}^{L_2}(\zeta_\beta \!- \!\tilde{\tilde{w}}_k) \over  \mu_{\tilde{\sigma}_{M_1+k}} -  \tilde{\tilde{w}}_k  }\right)\cr
&\& \quad  \times 
\det\pmatrix{  \ds{\mathop{K_{12}}^{\!N+L_2-M_2}}(\tilde{\tilde{z}}_k, \tilde{w}_j) & \ds{\mathop{K_{12}}^{\!N+L_2-M_2}}(\tilde{\tilde{z}}_k, \zeta_\beta) & P_{N+L_2-M_2}(\tilde{\tilde{z}}_k)
&\dots &P_{N+L_1 - M_1-1} (\tilde{\tilde{z}}_k )\cr
 \ds{\mathop{K_{12}}^{\!N+L_2-M_2}}(\xi_\alpha, \tilde{w}_j) & \ds{\mathop{K_{12}}^{\!N+L_2-M_2}}(\xi_\alpha, \zeta_\beta) & P_{N+L_2-M_2}(\xi_\alpha)
&\dots & P_{N+L_1 - M_1-1} (\xi_\alpha) \cr} \cr
&\&= {(-1)^{{1\over 2}(M_1+M_2)(M_1+M_2-1)}\prod_{n=N}^{N+L_2-M_2-1}\sqrt{h_n}
   \prod_{m=N}^{N+L_1 - M_1-1}\sqrt{h_n}\over 
   \Delta_{L_1}(\xi)\Delta_{L_2}(\zeta) \Delta_{M_1}(\eta)\Delta_{M_2}(\mu)}      \cr
&\& \quad \times \sum_{m=0}^{\min(M_1,M_2)} (-1)^m 
 \sum_{[\sigma,\tilde{\sigma}] }\sgn(\sigma) \sgn(\tilde{\sigma})
\prod_{i=1}^m\HH( \mu_{\tilde{\sigma}_i, \eta_{\sigma_i})}\cr
&\&  {\hskip - 20 pt} \times  
\det\pmatrix{ 
 \ds{\mathop{\KK_{21}}^{\!N+L_2-M_2}}(\mu_{\tilde{\sigma}_{M_1+k}}, \eta_{\sigma_{m+j}}) &{\hskip - 10 pt} \ds{\mathop{\KK_{22}}^{\!N+L_2-M_2}}(\mu_{\tilde{\sigma}_{M_1+k}}, \zeta_\beta) &{\hskip -5 pt} \tilde{\PP}_{N+L_2-M_2}(\mu_{\tilde{\sigma}_{M_1+k}})&{\hskip - 10 pt} 
\dots&{\hskip - 5 pt}\tilde{\PP}_{N+L_1 - M_1-1} (\mu_{\tilde{\sigma}_{M_1+k}} ) \cr
\ds{\mathop{\KK_{11}}^{\!N+L_2-M_2}}(\xi_\alpha, \eta_{\sigma_{m+j}}) & {\hskip - 10 pt}\ds{\mathop{K_{12}}^{\!N+L_2-M_2}}(\xi_\alpha, \zeta_\beta) & {\hskip - 10pt} P_{N+L_2-M_2}(\xi_\alpha) &{\hskip - 10 pt} 
\dots  &{\hskip - 20 pt} P_{N+L_1 - M_1-1} (\xi_\alpha) }
\cr
&\&
\cr
&\&
     \label{fullsigmasumint}
     \eea
     where
     \bea
    \ds{\mathop{\KK_{11}}^{{\hskip - 6 pt}J}}(\xi, \eta)&\& := 
        \int d\mu(x,y) {\prod_{\alpha=1}^{L_1} (\xi_\alpha -x) \over \eta - x}
           \  \ds{\mathop{K_{12}}^{{\hskip - 5 pt}J}}(\xi, y)
           \label{KK11}
        \\
    \ds{\mathop{\KK_{22}}^{{\hskip - 6 pt}J}}(\mu, \zeta)&\&:= 
      \int d\mu(x,y)  {\prod_{\beta=1}^{L_2} (\zeta_\beta -y) \over \mu - y}
           \  \ds{\mathop{K_{12}}^{{\hskip - 5 pt}J}}(x, \zeta)
             \label{KK22}
     \\
     \ds{\mathop{\KK_{21}}^{{\hskip - 6 pt} J}}(\mu,\eta) &\&:=
 \int d\mu(x,w) {\prod_{\alpha=1}^{L_1} (\xi_\alpha \! - \!x) \over \eta - x}
\! \int d\mu(z,y) {\prod_{\beta=1}^{L_2} (\zeta_\beta \!- \!y) \over \mu - y}
  \  \ds{\mathop{K_{12}}^{{\hskip - 5 pt}J}}(z,w) 
    \label{KK21}\cr
    &\&
  \\
       \ds{\HH} (\mu, \eta)&\&:=  \int d\mu(x,y){\prod_{\alpha=1}^{L_1} (\xi_\alpha -x)
       \prod_{\beta=1}^{L_2} (\zeta_\beta -y)   \over (\eta-x)(\mu-y)} 
         \label{HH}
    \\
  \tilde{\PP}_n(\mu) &\&:=\int d\mu(x,y)  {\prod_{\beta=1}^{L_2} (\zeta_\beta -y) \over \mu - y}
  P_n(x) .
  \label{PP}
             \eea
   Here, in the first line of eq. (\ref{fullsigmasumint}) we have cancelled the
   combinatorial factor $m!(M_1-m)!(M_2-m)!$ with the corresponding expression
   in the denominator of  $C^N_{m,M_1,M_2}$ and
   used the relation (\ref{partfn}) for ${\bf Z}_{N- M_1 - M_2 +m}^{(2)}$ to cancel the further factor $(N-M_1-M_2 +n)!$, leaving only the normalization factor
   $ \prod_{n=0}^{N+L_2-M_2-1}h_n
   \prod_{n=N+L_2-M_2}^{N+L_1 -M_1-1}\sqrt{h_n} $ in the numerator.
   Note also that, although the range of summation in $m$ is from $0$ to
   $\min(M_1, M_2)$, in the case when $ N <M_1 +M_2$, all terms with 
   $0 \le m <  M_1 + M_2 -N$ vanish, because the determinant factors,
   which are of dimension $(L_1 + M_2 - m) \times  (L_1 + M_2 -m)$, have entries
   that are formed from scalar products of vectors of dimension 
   $N +L_1-M_1 <   L_1 + M_2 -m$, and hence have less than maximal rank.

  We now note that the sum  over cosets  
   $[\sigma,\tilde{\sigma}]\in (S_{M_1} \times S_{M_2})/ S_{(m, M_1, M_2)}$
  in (\ref{fullsigmasumint}) is an expansion, as a sum over homogeneous polynomials in the terms $\{\HH(\mu_k, \eta_j)\}_{1\le j\le M_1 \atop1\le k \le M_2}$, 
  of the single  $(L_1 + M_2)\times (L_1 + M_2)$ determinant
  \bea
  {\bf I}^{(2)}  &\& =
  {  (-1)^{{1\over 2}(M_1+M_2)(M_1+M_2-1)} \prod_{n=N}^{N+L_2-M_2-1}\sqrt{h}_n
   \prod_{n=N}^{N+L_1-M_1-1}\sqrt{h_n}\over 
   \Delta_{L_1}(\xi)\Delta_{L_2}(\zeta) \Delta_{M_1}(\eta)\Delta_{M_2}(\mu)}   
 \cr
&\& {\hskip - 20 pt} \times 
\det\pmatrix{
 \ds{\mathop{\KK_{21}}^{\!N+L_2-M_2}}(\mu_k, \eta_j)
 - \ds{\HH} (\mu_k, \eta_j) &{\hskip - 10 pt} \ds{\mathop{\KK_{22}}^{\!N+L_2-M_2}}(\mu_k, \zeta_\beta) & \tilde{\PP}_{N+L_2-M_2}(\mu_k)&{\hskip - 10 pt} 
\dots&{\hskip - 5 pt}  \tilde{\PP}_{N+L_1 - M_1} (\mu_k )\cr
 \ds{\mathop{\KK_{11}}^{\!N+L_2-M_2}}(\xi_\alpha, \eta_j) & {\hskip - 10 pt}\ds{\mathop{K_{12}}^{\!N+L_2-M_2}}(\xi_\alpha, \zeta_\beta) & {\hskip - 5pt} P_{N+L_2-M_2}(\xi_\alpha) &{\hskip - 10 pt} 
\dots  &{\hskip - 10 pt} P_{N+L_1 - M_1} (\xi_\alpha) } \cr
 &\& =
  {  (-1)^{{1\over 2}(M_1+M_2)(M_1+M_2-1)}(-1)^{L_1 M_2} \prod_{n=N}^{N+L_2-M_2-1}h_n
   \prod_{n=N+L_2-M_2}^{N+L_1-M_1-1}\sqrt{h_n}\over 
   \Delta_{L_1}(\xi)\Delta_{L_2}(\zeta) \Delta_{M_1}(\eta)\Delta_{M_2}(\mu)}   
 \cr
&\&  {\hskip - 20 pt} \times
\det\pmatrix{
  \ds{\mathop{\KK_{11}}^{\!N+L_2-M_2}}(\xi_\alpha, \eta_j) & {\hskip - 10 pt}\ds{\mathop{K_{12}}^{\!N+L_2-M_2}}(\xi_\alpha, \zeta_\beta) & {\hskip - 10pt} P_{N+L_2-M_2}(\xi_\alpha) &{\hskip - 10 pt} 
\dots  &{\hskip - 10 pt} P_{N+L_1 - M_1} (\xi_\alpha)\cr
\ds{\mathop{\KK_{21}}^{\!N+L_2-M_2}}(\mu_k, \eta_j)
 - \ds{\HH} (\mu_k, \eta_j) &{\hskip - 10 pt} \ds{\mathop{\KK_{22}}^{\!N+L_2-M_2}}(\mu_k, \zeta_\beta) &{\hskip -5 pt} \tilde{\PP}_{N+L_2-M_2}(\mu_k)&{\hskip - 10 pt} 
\dots&{\hskip - 5 pt}  \tilde{\PP}_{N+L_1 - M_1} (\mu_k )}. \cr
&\&
\cr
&\&
\label{mainintegralcomplete}
\eea
Notice also that, by adding linear combinations of the last $L_1-M_1-L_2+M_2$
columns of the matrix in (\ref{mainintegralcomplete}}) the terms
\{$ \ds{\mathop{\KK_{11}}^{\!N+L_2-M_2}}(\xi_\alpha, \eta_j)$,
$\ds{\mathop{\KK_{21}}^{\!N+L_2-M_2}}(\mu_k, \eta_j)$\} may be replaced
by \{$\ds{\mathop{\KK_{11}}^{\!N+L_1-M_1}}(\xi_\alpha, \eta_j)$,
$\ds{\mathop{\KK_{21}}^{\!N+L_1-M_1}}(\mu_k, \eta_j)$\} to give
\bea
  {\bf I}^{(2)}  &\& =
  {  (-1)^{{1\over 2}(M_1+M_2)(M_1+M_2-1)} \prod_{n=N}^{N+L_2-M_2-1}\sqrt{h}_n
   \prod_{n=N}^{N+L_1-M_1-1}\sqrt{h_n}\over 
   \Delta_{L_1}(\xi)\Delta_{L_2}(\zeta) \Delta_{M_1}(\eta)\Delta_{M_2}(\mu)}   
 \cr
 &\&
{\hskip - 20 pt} \times
\det\pmatrix{
  \ds{\mathop{\KK_{11}}^{\!N+L_1-M_1}}(\xi_\alpha, \eta_j) & {\hskip - 10 pt}\ds{\mathop{K_{12}}^{\!N+L_2-M_2}}(\xi_\alpha, \zeta_\beta) & {\hskip - 10pt} P_{N+L_2-M_2}(\xi_\alpha) &{\hskip - 10 pt} 
\dots  &{\hskip - 10 pt} P_{N+L_1 - M_1} (\xi_\alpha)\cr
\ds{\mathop{\KK_{21}}^{\!N+L_1-M_1}}(\mu_k, \eta_j)
 - \ds{\HH} (\mu_k, \eta_j) &{\hskip - 10 pt} \ds{\mathop{\KK_{22}}^{\!N+L_2-M_2}}(\mu_k, \zeta_\beta) &{\hskip -5 pt} \tilde{\PP}_{N+L_2-M_2}(\mu_k)&{\hskip - 10 pt} 
\dots&{\hskip - 5 pt}  \tilde{\PP}_{N+L_1 - M_1} (\mu_k )}. \cr
&\&
\cr
&\&
\label{mainintegralcomplete1}
\eea
   As a final step we note that by separating the integrands in (\ref{KK11})--(\ref{PP}) into the sum of their principal parts at the poles 
   $x=\eta$ and $y=\mu $ and polynomial parts of degrees $\le L_1-1$ in $x$
   and $\le L_2-1$ in $y$, and using biorthogonality (which implies that $\ds{\mathop{K_{12}}^{\!N+L_2-M_2}}$
   is the kernel of an integral operator projecting onto the first $N+L_2-M_2$ biorthogonal  
   polynomials), the integrals (\ref{KK11})--(\ref{PP}), for $J= N+L_1-M_1$ or
   $J= N+L_2-M_2$  may be  reduced, at the values $\{\xi_\alpha, \zeta_\beta, \eta_j, \mu_k\}$, to the following
   \bea
   \ds{\mathop{\KK_{11}}^{\!N+L_1-M_1}}(\xi_\alpha, \eta_j)
   &\&= \prod_{\alpha=1}^{L_1}(\xi_\alpha - \eta_j)   \ds{\mathop{K_{11}}^{\!N+L_1-M_1}}(\xi_\alpha, \eta_j) \\
   \ds{\mathop{\KK_{22}}^{\!N+L_2-M_2}}(\mu_k, \zeta_\beta)
    &\&=   \prod_{\beta=1}^{L_2}(\zeta_\beta - \mu_k) \ds{\mathop{K_{22}}^{\!N+L_2-M_2}}(\mu_k, \zeta_\beta)  \\
    \ds{\mathop{\KK_{21}}^{\!N+L_1-M_1}}(\mu_k, \eta_j)
- \ds{\HH}(\mu_k, \eta_j)
     &\& =  \prod_{\alpha=1}^{L_1}(\xi_\alpha - \eta_j)
       \prod_{\beta=1}^{L_2}(\zeta_\beta - \mu_k) 
   \ds{\mathop{K_{21}}^{\!N+L_1-M_1}}(\mu_k, \eta_j) \\
  \tilde{\PP}_n(\mu_k) 
&\&=   \prod_{\beta=1}^{L_2}(\zeta_\beta - \mu_k)  \tilde{P}_n(\mu_k), \quad n \ge L_2 .
    \eea
Substituting these expressions into the determinant  term in (\ref{mainintegralcomplete1}), factoring out the diagonal matrices  $\diag(1, \dots ,1 ,\prod_{\beta=1}^{L_2}(\zeta_\beta - \mu_1), \dots , \prod_{\beta=1}^{L_2}(\zeta_\beta - \mu_{M_2}))$,
making the replacements $ \ds{\mathop{K_{11}}^{\!N+L_1-M_1}}(\xi_\alpha, \eta_j),\ds{\mathop{K_{21}}^{\!N+L_1-M_1}}(\mu_k, \eta_j) \ra  \ds{\mathop{K_{11}}^{\!N+L_2-M_2}}(\xi_\alpha, \eta_j), \ds{\mathop{K_{21}}^{\!N+L_2-M_2}}(\mu_k, \eta_j)$
in the first $L_2 +M_1$ columns of the determinant,  by adding linear combinations of the last $L_1-M_1-L_2+M_2$ columns
 and factoring our the diagonal matrices $\diag(\prod_{\alpha=1}^{L_1}(\xi_\alpha - \eta_1), \dots \prod_{\alpha=1}^{L_1}(\xi_\alpha - \eta_{M_1}), 1 \dots , 1) )$ on the left and right when evaluating the determinant then gives the relation (\ref{mainintegral}).

\br
Although the computation was done here for the case when $N \ge M_1, 
N \ge M_2 $, the other cases, in which $M_1 -L_1 \le N \le M_1$, or 
$M_2 -L_2 \le N \le M_2$, or both, may be derived similarly, leading to the
same formula (\ref{mainintegral}).
\er

   Formulae analogous to  (\ref{singleintegralMgN}) for the cases
   $N+L_1-M_1 \ge 0, N+L_2-M_2 \le 0$ and $N+L_1-M_1 \le 0, N+L_2-M_2 \le 0$
   may similarly be deduced using the second form of the partial fraction identity
   (\ref{partialfrac2})  whenever the inequalities relating the degrees of 
   the numerator and denominator polynomials in (\ref{I2_intermediate}) require it.
   These are derived in the next two subsections.
   
\subsection {Case 2. $N + L_1 - M_1  \ge 0 \ge  N +L_2 -M_2 $}

In this case the integral  ${\bf I}^{(2)}_N$  is given by the following determinantal
expression:
\bea
 {\bf I}^{(2)}_N &\&=
 \tilde{\epsilon}(L_1,L_2,M_1,M_2) 
 {\prod_{n=0}^{{}_{N-M_1+L_1-1}}{\hskip -6 pt}\sqrt{h_n} 
 \prod_{n=0}^{{}_{M_2-N-L_2-1}}{\hskip -6 pt}\sqrt{h_n} \over
 \prod_{n=0}^{N-1} h_n} \cr
 &\& \quad \times
 {  \prod_{\alpha=1}^{L_1} \prod_{j=1}^{M_2} (\xi_\alpha- \eta_j) \prod_{\beta=1}^{L_2}\prod_{k=1}^{M_2} (\zeta_\beta - \mu_k)\over
\Delta_{L_1}(\xi) \Delta_{L_2}(\zeta)\Delta_{M_1}(\eta)\Delta_{M_2}(\mu) } 
\det(G),
\label{mainintegral2}
\eea
where
\be
 \tilde{\epsilon}(L_1,L_2,M_1,M_2):=(-1)^{{1\over 2}N(N-1) +{1\over 2}M_1(M_1+1) 
 + {1\over 2} L_2(L_2-1) +M_1 N + M_2L_1 + M_1 L_2 + L_1 L_2}
\ee
and $G$ is the $(L_1+M_2) \times (L_1+M_2)$ matrix
\be 
G:=  \det\pmatrix{{1\over  \eta_j - \xi_\alpha }& 0 &
 P_b(\xi_\alpha) & 0   \cr
H(\mu_k, \eta_j)& {1\over \mu_k - \zeta_\beta} & \tilde{P}_b(\mu_k)  & S_m(\mu_k) }.
\label{Gmatrix2}
\ee
Here the row indices are, sequentially, $1\le \alpha \le L_1$ and $1\le k \le M_2$,
and the column indices $1\le j \le M_1$, $1\le \beta \le L_2$, $ 0\le b \le N-M_1+ L_1-1$
and $0\le m \le M_2 - N - L_2 -1 $. Note that  the first three column blocks of (\ref{Gmatrix2}) coincide with those of  the matrix $G$ defined in (\ref{Gmatrix}) if one understands the orthogonal polynomials $P_n(x), S_n(y)$ and their Hilbert transforms to vanish for negative $n$. Formula (\ref{mainintegral2}) is valid whenever
$N + L_1 - M_1 \ge 0$, $ N +L_2 -M_2 \le 0$,
but it is easier to demonstrate assuming the stronger conditions
\be
N\ge M_1, \quad  N +L_2 -M_2 \le 0,
\ee
 which is what we do in the following. The intermediate case,
 when $M_1 -L_1 \le N \le M_1$ may be proved through a similar computation.

Expressing ${\bf I}^{(2)}_N $ in this case as:
\bea
 {\bf I}^{(2)}_N &\&:= {1\over {\bf Z}^{(2)}_N}\prod_{a=1}^N\left( \int d\mu(x_a,y_a)
 {\prod_{\alpha=1}^{L_1}(\xi_\alpha - x_a) \over \prod_{j=1}^{M_1}(\eta_j - x_a)  \prod_{k=1}^{M_2}(\mu_k - y_a)} \right)\cr
&\&{\hskip 60 pt}\times { \Delta_N(x)\Delta_{N+L_2}(y,\zeta) \over \Delta_{L_2}(\zeta)}
\eea
and applying identity (\ref{partialfrac1}) with respect to the $(x,\eta)$ variables 
and (\ref{partialfrac2}) with respect to the $((y,\zeta)),\mu)$ variables gives
\bea
 {\bf I}^{(2)}_N &\&= {(-1)^{M_1N +{1\over 2} (N+L_2)(N+L_2-1)}
\prod_{\beta=1}^{L_2}  \prod_{k=1}^{M_2}(\mu_k - \zeta_\beta)\over
 N!  (M_2-N-L_2)!(\prod_{n=0}^{N-1}h_n) \Delta_{L_2}(\zeta)\Delta_{M_1}(\eta)\Delta_{M_2}(\mu)} 
\cr
 &\&\quad \times
 \sum_{a_1\le \dots \le a_{M_1}}{\hskip -10 pt}(-1)^{\sum_{j=1}^{M_1} a_j}\sum_{\sigma\in S_{M_1}}{\hskip -5 pt}
 \sgn(\sigma){\hskip -5 pt}\sum_{\wt{\sigma}\in S_{M_2}}{\hskip -5 pt}\sgn(\wt{\sigma})
 {\Delta_{M_2-L_2-N}(\eta_{\wt{\sigma}_{N+L_2+1}}, \dots,
  \eta_{\wt{\sigma}_{M_2}})\over
  \prod_{\beta=1}^{L_2} ( \mu_{\wt{\sigma}_{N+\beta}} - \zeta_\beta)}
 \cr
 &\& \quad \times
 \left(\prod_{a=1}^N \int {d\mu(x_a, y_a) \prod_{\alpha=1}^{L_1}(\xi_\alpha - x_a)
 \over \mu_{\wt{\sigma}_a} - y_a}
\right)  
 {\Delta_{N-M_1}(x[a]) \over \prod_{j=1}^{M_1}(\eta_{\sigma_j} - x_j)}
 \cr
 &\& =  {(-1)^{M_1N +{1\over 2} (N+L_2)(N+L_2-1)+{1\over 2} M_1(M_1+1)}
 \prod_{k=1}^{L_2}\prod_{\beta=1}^{M_2} (\mu_k - \zeta_\beta)\over
 (M_2-N-L_2)!M_1!(N-M_1)! (\prod_{n=0}^{N-1} h_n)
\Delta_{L_1}(\xi) \Delta_{L_2}(\zeta)\Delta_{M_1}(\eta)\Delta_{M_2}(\mu) } \cr
&\& \quad \times \sum_{\sigma\in S_{M_1}}{\hskip -5 pt}
 \sgn(\sigma){\hskip -5 pt}\sum_{\wt{\sigma}\in S_{M_2}}{\hskip -5 pt}\sgn(\wt{\sigma})
 \prod_{j=1}^{M_1}
  \left(\int {d\mu(z_j, w_j) \prod_{\alpha=1}^{L_2} (\xi_\alpha - z_j)
  \over (\eta_{\sigma_j} - z_j)(\mu_{\wt{\sigma}_j} - w_j)} \right) 
  \cr
  &\& \quad \times \prod_{a=1}^{N-M_1}\left(\int {d\mu(x_a, y_a)
  \over \mu_{\wt{\sigma}_{M_1+a}} - y_a} \right) 
  \det\pmatrix{ x_a^b \cr \xi_\alpha^b}_{1\le a \le N-M_1, \ 1\le \alpha \le L_1
  \atop 0 \le b \le N-M_1+L_1-1}\cr
    &\& \quad \times
     { \det(\mu^m_{\wt{\sigma}_{N+L_2+k}})_{{}_{0\le m \le M_2-N-L_2-1, \ 1\le k \le M_2-N-L_2}}
    \over \prod_{\beta=1}^{L_2} (\mu_{\wt{\sigma}_{N+\beta}}-\zeta_\beta)}\\
    &\&= {(-1)^{M_1N +{1\over 2} (N+L_2)(N+L_2-1)+{1\over 2} M_1(M_1+1)}
 \prod_{k=1}^{L_2}\prod_{\beta=1}^{M_2} (\mu_k - \zeta_\beta)\over
 (M_2-N-L_2)!M_1!(N-M_1)! (\prod_{n=0}^{N-1} h_n)
\Delta_{L_1}(\xi) \Delta_{L_2}(\zeta)\Delta_{M_1}(\eta)\Delta_{M_2}(\mu) } \cr
&\& \quad \times \sum_{\sigma\in S_{M_1}}{\hskip -5 pt}
 \sgn(\sigma){\hskip -5 pt}\sum_{\wt{\sigma}\in S_{M_2}}{\hskip -5 pt}\sgn(\wt{\sigma})
 \prod_{j=1}^{M_1}\left(\HH^\xi(\mu_{\wt{\sigma}_j}, \eta_{\sigma_j}\right)
 \det\pmatrix{ X_b(\mu_{\wt{\sigma}_{M_1+a}}) \cr \xi_\alpha^b}_{1\le a \le N-M_1, \ 1\le \alpha \le L_1 \atop 0 \le b \le N-M_1+L_1-1}\cr
&\& \quad \times
 \prod_{\beta=1}^{L_2}\left({1\over  \mu_{\wt{\sigma}_{N+\beta}}-\zeta_\beta}\right)
  \det(\mu^m_{\wt{\sigma}_{N+L_2+k}})_{{}_{0\le m \le M_2-N-L_2-1, \ 1\le k \le M_2-N-L_2}},
  \label{detprod2}
\eea
where
\bea
\HH^\xi(\mu, \eta) &\&:= 
\int {d\mu(x,y)\prod_{\alpha=1}^{L_1} (\xi_\alpha- x)\over (\eta-x)(\mu-y)}
\label{Hxi} \\
X_b(\mu) &\&:= \int d\mu(x,y) {x^b \over \mu - y}.
\eea
We now note that (\ref{detprod2}) is just the block determinant expansion
of
\bea
 {\bf I}^{(2)}_N &\&=
 {(-1)^{M_1N +{1\over 2} (N+L_2)(N+L_2-1)+{1\over 2} M_1(M_1+1)}
 \prod_{k=1}^{L_2}\prod_{\beta=1}^{M_2} (\mu_k - \zeta_\beta)\over
(\prod_{n=0}^{N-1} h_n)
\Delta_{L_1}(\xi) \Delta_{L_2}(\zeta)\Delta_{M_1}(\eta)\Delta_{M_2}(\mu) } \cr
&\& \quad \times
\det\pmatrix{ \HH^\xi(\mu_k, \eta_j) & X_b(\mu_k) & {1\over \mu_k - \zeta_\beta} & \mu_k^m \cr
0 & \xi_\alpha^b & 0 & 0 }_{ 0\le b \le N-M_1-L_1-1, \  0\le m \le M_2 - N - L_2 -1 \atop 1\le \alpha \le L_1, \ 1\le \beta \le L_2, \ 1\le j \le M_1, \ 1\le k \le M_2}
\label{det2intermediate}
\eea
(The combinatorial factors $(M_2-L-L_2)!M_1!(N-M_1)!$ in (\ref{detprod2}) are cancelled by the multiplicity with which the subdeterminant factors occur
in the sums over $\sigma \in S_{M_1}$, $\wt{\sigma}\in S_{M_2}$.) 
A further simplification can be achieved by applying elementary column operations.
To see this, we take the integrations appearing in the matrix elements of
(\ref{det2intermediate}) outside the determinant, re-writing it as:
\bea
 {\bf I}^{(2)}_N &\&=
 {(-1)^{M_1N +{1\over 2} (N+L_2)(N+L_2-1)+{1\over 2} M_1(M_1+1)}
 \prod_{k=1}^{L_2}\prod_{\beta=1}^{M_2} (\mu_k - \zeta_\beta)\over
(\prod_{n=0}^{N-1} h_n)
\Delta_{L_1}(\xi) \Delta_{L_2}(\zeta)\Delta_{M_1}(\eta)\Delta_{M_2}(\mu) } \cr
&\& \quad \times 
\prod_{k=1}^{M_2}\left(\int {d\mu(x_k, y_k)\over (\mu_k - y_k)}\right)
\det\pmatrix{{ \prod_{\alpha=1}^{L_1}(\xi_\alpha- x_k)\over (\eta_j-x_k)} &  x_k^b & {1\over \mu_k - \zeta_\beta} & \mu_k^m \cr
0 & \xi_\alpha^b & 0 & 0 }.
\label{det2inter2}
\eea
Now, viewing ${ \prod_{\alpha=1}^{L_1}(\xi_\alpha- x_k)\over (\eta_j-x_k)}$ as a rational function in $x_k$ with a simple pole at  $x_k=\eta_j$, we may re-express it as
the sum of the pole term plus a polynomial of degree $\le L_1-1$
\be
{ \prod_{\alpha=1}^{L_1}(\xi_\alpha- x_k)\over (\eta_j-x_k)}
= { \prod_{\alpha=1}^{L_1}(\xi_\alpha- \eta_j)\over (\eta_j-x_k)} 
+ \sum_{\gamma=0}^{L_1-1} \Lambda_{j \gamma} x_k^\gamma,
\ee
where
\be
\sum_{\gamma=0}^{L_1-1} \Lambda_{j \gamma} \xi_\alpha^\gamma
=\prod_{\beta= 1 \atop\beta\neq \alpha}^{L_1} (\xi_\beta - \eta_j).
\ee
Since $N - M_1+L_1-1\ge L_1-1$, all monomials in $x_k$ of degree  $\le L_1-1$ appear in the second column block of the determinant in ({\ref{det2inter2}).   We may 
therefore add linear combinations of these columns  to those in the first block to obtain  the equivalent expression
\bea &\&
\det\pmatrix{{ \prod_{\alpha=1}^{L_1}(\xi_\alpha- x_k)\over (\eta_j-x_k)} &  x_k^b & {1\over \mu_k - \zeta_\beta} & \mu_k^m \cr
0 & \xi_\alpha^b & 0 & 0 }
= \det\pmatrix{{ \prod_{\alpha=1}^{L_1}(\xi_\alpha- \eta_j)\over (\eta_j-x_k)} &  x_k^b & {1\over \mu_k - \zeta_\beta} & \mu_k^m \cr
-\prod_{\beta= 1 \atop\beta\neq \alpha}^{L_1} (\xi_\beta - \eta_j) & \xi_\alpha^b & 0 & 0 }
\cr
&\& = \prod_{j=1}^{{}_{M_2}}\prod_{\alpha=1}^{{}_{L_1}} (\xi_\alpha- \eta_j)\ 
 \det\pmatrix{{ 1\over \eta_j-x_k} &  x_k^b & {1\over \mu_k - \zeta_\beta} & \mu_k^m \cr
{1\over \eta_j - \xi_\alpha}  & \xi_\alpha^b & 0 & 0 } \cr
&\& = \prod_{n=0}^{{}_{N-M_1+L_1-1}}{\hskip -16 pt}\sqrt{h_n} \prod_{n=0}^{{}_{M_2-N-L_2-1}}{\hskip -16 pt}\sqrt{h_n}
\prod_{\alpha=1}^{L_1} \prod_{j=1}^{M_2} (\xi_\alpha- \eta_j)\  \det\pmatrix{{ 1\over \eta_j-x_k} &  P_b(x_k) & {1\over \mu_k - \zeta_\beta} & S_m(\mu_k) \cr
{1\over \eta_j - \xi_\alpha}  & P_b(\xi_\alpha) & 0 & 0 }, \cr
&\&
\eea
where further elementary column operations were made in the last lign to replace the
monomials  $x^b_k, \xi^b_\alpha$ and $\mu^m_k$ by the corresponding biorthogonal polynomials. Substituting this expression into (\ref{det2inter2}) and evaluating the integrals gives
\bea
 {\bf I}^{(2)}_N &\&=
 {(-1)^{M_1N +{1\over 2} (N+L_2)(N+L_2-1)+{1\over 2} M_1(M_1+1)}
\prod_{\alpha=1}^{L_1} \prod_{j=1}^{M_2} (\xi_\alpha- \eta_j) \prod_{k=1}^{L_2}\prod_{\beta=1}^{M_2} (\mu_k - \zeta_\beta)\over
(\prod_{n=0}^{N-1} h_n)
\Delta_{L_1}(\xi) \Delta_{L_2}(\zeta)\Delta_{M_1}(\eta)\Delta_{M_2}(\mu) } \cr
&\& \quad \times 
\prod_{n=0}^{{}_{N-M_1+L_1-1}}{\hskip -16 pt}\sqrt{h_n} \prod_{n=0}^{{}_{M_2-N-L_2-1}}{\hskip -16 pt}\sqrt{h_n}\ 
 \det\pmatrix{H(\mu_k, \eta_j) & \tilde{P}_b(\mu_k) & {1\over \mu_k - \zeta_\beta} & S_m(\mu_k) \cr
{1\over \eta_j - \xi_\alpha}& P_b(\xi_\alpha) & 0 & 0 }.
\label{mainintegral2inter}
\eea
Finally, reordering the rows and columns suitably, bringing the third column block
into the second position and interchanging the two row blocks, we arrive
at the expression (\ref{mainintegral2}).

\subsection {Case 3.  $N + L_1 - M_1 \le 0$, $N +L_2 - M_2 \le 0$}
In this case the integral  ${\bf I}^{(2)}_N$  is given by the following determinantal
expression:
\bea
 {\bf I}^{(2)}_N &\&=
 {\prod_{n=0}^{{}_{M_1-L_1-N}}{\hskip -6 pt}\sqrt{h_n} 
 \prod_{n=0}^{{}_{M_2-L_2-N}}{\hskip -6 pt}\sqrt{h_n} \over
 \prod_{n=0}^{N-1} h_n} \cr
 &\& \quad \times
 {  \prod_{\alpha=1}^{L_1} \prod_{j=1}^{M_2} (\xi_\alpha- \eta_j) \prod_{\beta=1}^{L_2}\prod_{k=1}^{M_2} (\zeta_\beta - \mu_k)\over
\Delta_{L_1}(\xi) \Delta_{L_2}(\zeta)\Delta_{M_1}(\eta)\Delta_{M_2}(\mu) } 
\det(G),
\label{mainintegral3}
\eea
where $G$ is the $(M_1+M_2 -N) \times (M_1+M_2 -N)$ matrix
\be 
G:= \det\pmatrix{ H(\mu_k, \eta_j) & S_m(\mu_k) & {1\over \mu_k - \zeta_\beta}  \cr
P_\ell(\eta_j) & 0 & 0  \cr
 {1\over  \eta_j - \xi_\alpha}& 0 & 0 },
\label{Gmatrix3}
\ee
with rows labelled consecutively by $1\le \alpha \le L_1$,  $1\le k \le M_2$,
 and $0\le \ell \le M_1 - L_1 -N$ and the columns by 
 $1\le j \le M_1$, $1\le \beta \le L_2$ and $0\le m \le M_2 - L_2 -N$.

  To derive this formula, we begin by expressing the integral in the form
\bea
 {\bf I}^{(2)}_N &\&= {\prod_{\alpha=1}^{L_1}\prod_{j=1}^{M_1} (\xi_\alpha -\eta_j)\prod_{\beta=1}^{L_2}\prod_{k=1}^{M_2}
 ( \zeta_\beta -\mu_k)\over Z_M^{(2)} 
\Delta_{L_1}(\xi) \Delta_{L_2}(\zeta)} 
 \int d\mu(x_1, y_1)  \dots \int d\mu(x_N, y_N)
 \cr
  &\&  \quad \times  {\Delta_{N+L_1}(x, \xi)
 \Delta_{N+L_2}(y, \eta) \over\prod_{a=1}^N\prod_{\alpha=1}^{L_1}\prod_{j=1}^{M_1} (\xi_\alpha -\eta_j) (\eta_j -x_a)\prod_{\beta=1}^{L_2}\prod_{k=1}^{M_2}
 ( \zeta_\beta -\mu_k)  (\mu_k - y_a)}. 
 \eea
  Using the identity (\ref{partialfrac2}) twice, first with the $x_a$ variables replaced by the combined set $(x_a,\xi_j)$  and $N\ra N+L_1$, and also for the set $(y_a,\zeta_k)$ with $N \ra N+L_2$, we obtain
 \bea
 {\bf I}^{(2)}_N &\&= {(-1)^{L_1(N+M_1)+L_2(N+M_2) +{1\over 2}L_1(L_1+1) +{1\over 2}L_2(L_2+1)}\prod_{\alpha=1}^{L_1}\prod_{j=1}^{M_1} (\xi_\alpha -\eta_j)\prod_{\beta=1}^{L_2}\prod_{k=1}^{M_2}
 ( \zeta_\beta -\mu_k)\over N!(M_1-L_1-N)!(M_2-L_2-N)!(\prod_{n=0}^{N-1}{h_n})\ 
\Delta_{L_1}(\xi) \Delta_{L_2}(\zeta) \Delta_{M_1}(\eta) \Delta_{m_2}(\mu)} \cr
&\& \quad \times \sum_{\sigma\in S_{M_1}}\sum_{\wt{\sigma}\in S_{M-2}}
{\Delta_{M_1-L_1-N}(\eta_{\sigma_{N+L_1+1}}, \dots, \eta_{\sigma_{M_1}})
\Delta_{M_2-L_2-N}(\mu_{\wt{\sigma}_{N+L_2+1}}, \dots, \mu_{\wt{\sigma}_{M_2}})
\over \prod_{\alpha=1}^{L_1}(\eta_{\sigma_{N+\alpha}}- \xi_\alpha)
\prod_{\beta=1}^{L_2}(\mu_{\wt{\sigma}_{N+\beta}}- \zeta_\beta)}
\cr
\&\& \quad \times 
\prod_{a=1}^N\left( \int {d\mu(x_a, y_a)
\over (\eta_{\sigma_a} - x_a) (\mu_{\wt{\sigma}_a} - y_a)}\right)
 \cr
&\&= {(-1)^{L_1(N+M_1)+L_2(N+M_2) +{1\over 2}L_1(L_1+1) +{1\over 2}L_2(L_2+1)}\prod_{\alpha=1}^{L_1}\prod_{j=1}^{M_1} (\xi_\alpha -\eta_j)\prod_{\beta=1}^{L_2}\prod_{k=1}^{M_2}
 ( \zeta_\beta -\mu_k)\over N!(M_1-L_1-N)!(M_2-L_2-N)!(\prod_{n=0}^{N-1}{h_n})\ 
\Delta_{L_1}(\xi) \Delta_{L_2}(\zeta) \Delta_{M_1}(\eta) \Delta_{m_2}(\mu)} \cr
&\& \quad \times \sum_{\sigma\in S_{M_1}}\sum_{\wt{\sigma}\in S_{M-2}}
\left( \prod_{a=1}^N  H(\mu_{\wt{\sigma}_a},\eta_{\sigma_a}) \right)
\left( \prod_{\alpha=1}^{L_1}{1\over \eta_{\sigma_{N+\alpha}}- \xi_\alpha}\right)
\left( \prod_{\beta=1}^{L_2}{1\over \mu_{\wt{\sigma}_{N+\beta}}- \zeta_\beta}\right)
\cr
&\& \quad \times
\det(\eta^\ell_{\sigma_{N+L_1+j}})_{1\le j \le M_1-L_1 -N\atop 0\le \ell \le  M_1-L_1 -N}
\ \det(\mu^m_{\wt{\sigma}_{N+L_2+k}})_{1\le k \le M_2-L_2 -N\atop 0\le m \le  M_2-L_2 -N}
\cr
\&\& = {(-1)^{L_1(N+M_1)+L_2(N+M_2) +{1\over 2}L_1(L_1+1) +{1\over 2}L_2(L_2+1)}\prod_{\alpha=1}^{L_1}\prod_{j=1}^{M_1} (\xi_\alpha -\eta_j)\prod_{\beta=1}^{L_2}\prod_{k=1}^{M_2}
 ( \zeta_\beta -\mu_k)\over (\prod_{n=0}^{N-1}{h_n})\ 
\Delta_{L_1}(\xi) \Delta_{L_2}(\zeta) \Delta_{M_1}(\eta) \Delta_{m_2}(\mu)} \cr
&\& \quad \times \sum_{\sigma\in S_{M_1}}\sum_{\wt{\sigma}\in S_{M-2}}
\left( \prod_{a=1}^N  H(\mu_{\wt{\sigma}_a},\eta_{\sigma_a}) \right)
\left( \prod_{\alpha=1}^{L_1}{1\over \eta_{\sigma_{N+\alpha}}- \xi_\alpha}\right)
\left( \prod_{\beta=1}^{L_2}{1\over \mu_{\wt{\sigma}_{N+\beta}}- \zeta_\beta}\right)
\cr
&\& \quad \times
\det\pmatrix{ H(\mu_k, \eta_j) & {1\over \mu_k - \zeta_\beta} & \mu_k^m \cr
{1\over \eta_j - \xi_\alpha} & 0 & 0 \cr
\eta_j^\ell & 0 & 0 },
 \eea  
 where the rows are labelled consecutively by $1\le k \le M_2$, $1\le \alpha \le L_1$
 and $0\le \ell \le M_1 - L_1 -N$ and the columns by 
 $1\le j \le M_1$, $1\le \beta \le L_2$
 and $0\le m \le M_2 - L_2 -N$. 
Finally,  interchanging the two row blocks, and replacing the monomial entries
$\eta_j^l$ and $\mu_k^m$ by the biorthogonal polynomials $P_j(\eta_j)k)$
and $S_m(\mu_k)$ respectively, with suitably modified normalization
factors, we arrive at the expression (\ref{mainintegral3}) for ${\bf I}^{(2)}_N$.



\begin{thebibliography}{99}

\bibitem{AV} G. Akemann and G. Vernizzi, ``Characteristic polynomials of complex random matrix models'', {\it Nucl. Phys.} {\bf B 660}, 532--556 (2003).

\bibitem{Be} M. Berg\`ere, ``Biorthogonal polynomials for potentials of two variables and external sources at the denominator'', hep-th/0404126

\bibitem{BEH1} M. Bertola, B. Eynard and J. Harnad, ``Duality,
Biorthogonal Polynomials and Multi--Matrix Models'',
{\it Commun. Math. Phys.} {\bf 229}, 73--120 (2002).

\bibitem{BEH2} M. Bertola, B. Eynard and J. Harnad,
``Differential systems for biorthogonal polynomials appearing in
2-matrix models and the associated Riemann-Hilbert problem'',
{\it Commun. Math. Phys.} {\bf 243}, 193--240 (2003).

\bibitem{BDS} J. Baik, P. Deift and E. Strahov, ``Products and ratios of characteristic polynomials of random Hermitian matrices'', J. Math. Phys. {\bf 44}, 3657--3670 (2003). 
	 
\bibitem{BH} E. Brezin and S. Hikami, ``Characteristics polynomials of random matrices'',  {\it Commun. Math. Phys.} {\bf 214}, 111--135 (2000).

\bibitem{EM} B. Eynard  and M.L. Mehta, ``Matrices coupled in a chain:
eigenvalue correlations'', {\em J. Phys. A} {\bf 31}, 4449--4456 (1998).

\bibitem{FS} Y.V.  Fyodorov and E. Strahov, ``An exact formula for spectral correlation functions of random matrices'', {\it J. Phys. A}  {\bf 36}, 3203--3213 (2003).

\bibitem{GMO} A. Gerasimov, A. Marshakov, A. Mironov, A. Morozov and A.Yu.Orlov, 
``Matrix Models of 2D Gravity and Toda Theory'', {\it Nucl. Phys. } {\bf B357}, 565--618
(1991). 

\bibitem{G} Y. L. Geronimus, ``Orthogonal Polynomials: Estimates, asymptotic formulas, and series of polynomials orthogonal on the unit circle and on an interval'', Consultants Bureau, New York (1961).

\bibitem{HO1} J. Harnad and A. Yu. Orlov, `` Fermionic construction of partition functions for two-matrix models and perturbative Schur function expansions'',  CRM preprint-3195 (2005), math-ph/0512056.  {\it J. Phys. A} Special issue on ``Random matrices, random processes and integrable systems'' (to appear, July - Aug., 2006)

\bibitem{HO2} J. Harnad and A. Yu. Orlov, ``Fermionic approach to the evaluation of integrals of rational symmetric functions'',  CRM preprint (2006). 

\bibitem{IZ}  C. Itzykson and J.-B. Zuber,  ``The planar approximation. II'', {\it J. Math. Phys.}  {\bf 21} 411--421 (1980).

\bibitem{Sz}  G. Szeg\"o, ``Orthogonal Polynomials'', {\it American Mathematical Society Publications} Vol. {\bf 23}.A.M.S.. Providence, R.I., 4th ed. (1975).

\bibitem{U} V.B. Uvarov, ``The connection between systems of polynomials orthogonal with respect to different distribution functions'', U.S.S.R. Comput. Math. and Math. Phys. {\bf 9}, 25--36 (1969).




\end{thebibliography}
\end{document}